\documentclass[acmsmall,screen,nonacm]{acmart}

\usepackage{booktabs}
\usepackage{tabularx}
\usepackage{longtable,afterpage}
\usepackage{microtype}
\usepackage{flafter}

\setcopyright{none}
\title{Characterizing and Detecting Unfaithful Work Reports by Coding Agents}
\makeatletter
\renewcommand{\@ACM@checkaffil}{}
\makeatother

\author{Wenhao Yang}
\email{yangwh@stu.pku.edu.cn}
\affiliation{\institution{Peking University}\city{Beijing}\country{China}}

\author{Runzhi He}
\email{rzhe@pku.edu.cn}
\affiliation{\institution{Peking University}\city{Beijing}\country{China}}

\author{Minghui Zhou}
\authornote{Corresponding authors.}
\email{zhmh@pku.edu.cn}
\affiliation{\institution{Peking University}\city{Beijing}\country{China}}

\author{Pengfei Gao}
\authornotemark[1]
\email{gaopengfei@microsoft.com}
\affiliation{\institution{Microsoft}}

\author{Dayi Lin}
\email{dayi.lin@microsoft.com}
\affiliation{\institution{Microsoft}}

\author{Qingwei Lin}
\email{qlin@microsoft.com}
\affiliation{\institution{Microsoft}}

\author{Dongmei Zhang}
\email{dongmeiz@microsoft.com}
\affiliation{\institution{Microsoft}}

\renewcommand{\shortauthors}{Yang et al.}

\newcommand{\NaturalPositiveReports}{162}
\newcommand{\NaturalPositiveShare}{37.24}

\newcommand{\EvaluationTargets}{2356}
\newcommand{\EvaluationReports}{409}

\newcommand{\NaturalDetectorMacro}{93.06}
\newcommand{\NaturalBaselineMacro}{80.42}
\newcommand{\NaturalDetectorPrecision}{90.93}
\newcommand{\NaturalDetectorRecall}{86.98}
\newcommand{\NaturalBaselinePrecision}{78.75}
\newcommand{\NaturalBaselineRecall}{59.71}
\newcommand{\NaturalMacroDelta}{12.64}
\newcommand{\NaturalDeltaLow}{9.71}
\newcommand{\NaturalDeltaHigh}{15.49}

\newcommand{\NaturalDetectorFaithfulPrecision}{96.67}
\newcommand{\NaturalDetectorFaithfulRecall}{97.76}
\newcommand{\NaturalDetectorFaithfulFone}{97.21}
\newcommand{\NaturalDetectorUnfaithfulFone}{88.91}
\newcommand{\NaturalBaselineFaithfulPrecision}{90.20}
\newcommand{\NaturalBaselineFaithfulRecall}{95.83}
\newcommand{\NaturalBaselineFaithfulFone}{92.93}
\newcommand{\NaturalBaselineUnfaithfulFone}{67.92}

\begin{document}
\raggedbottom
\begin{abstract}
Coding agents increasingly carry out long-horizon development tasks, autonomously planning and executing multi-step actions that span files, tools, and sessions. Because developers rarely follow each action, and agents resuming the work depend on records of what was done, the agent's work report often becomes the primary account of what was changed and established. When the report contains false or unsupported claims, it gives recipients false assurance: incomplete work may be accepted without verification, incorrect assumptions about the current state can propagate to later steps and agents, and evaluators that credit stated success may reward the report rather than the work. Yet little is known about what unfaithful reporting looks like in real development work, or how developers and agents handle it once it surfaces. Starting from real developer--agent conversations in SWE-chat, we conduct a systematic mixed-methods analysis of self-admitted unfaithful reports. We identify eleven subtypes in four categories: misreported observations, misattributed evidence, overstated task success, and faulty technical explanations, with overstated task success the most common. Handling these problems involves on average 4.3 further user turns and 12.71 minutes of recorded response and execution time, and users express dissatisfaction more often. Among 435 work reports that we separately sample and annotate, \NaturalPositiveShare{}\% contain at least one unfaithful claim. We also develop FaithDetector, a detector for unfaithful reporting informed by the observed patterns. On naturally sampled reports, FaithDetector attains \NaturalDetectorMacro{}\% macro-F1, about 13 percentage points above the baseline. Together, our findings and FaithDetector provide empirical evidence and an automated method for checking reporting faithfulness, informing agent evaluation and training feedback.

\end{abstract}

\maketitle

\section{Introduction}\label{sec:introduction}

Coding agents now undertake long-horizon development tasks, autonomously planning and executing multi-step actions that edit files, run commands, and change runtime state \cite{bouzenia2025trajectories,tang2026coding-agents-fail}. As agents assume more of this work, the developer's role shifts from writing every change toward delegating, orchestrating, and supervising it \cite{fawzy2026vibe-practice}. Vibe coding, a widely adopted practice, makes this shift explicit: developers describe goals in natural language and rely on minimal review of the generated code \cite{fawzy2026vibe-practice}. In asynchronous workflows, developers likewise hand off a task and return to a report rather than following each intermediate action \cite{willison2025async-code-research}, and they consult progress updates and change summaries to decide what to review, accept, or investigate further \cite{tang2026programming-chat,dhanorkar2026oversight}. When the work spans several sessions, agents that resume it in a fresh context depend on records of what has already been done \cite{young2025long-running-harnesses}. The agent's report thereby becomes the primary account of what has changed and what has been established, and the basis on which developers and subsequent agents decide how to proceed.

\begin{figure}[t]
\centering
\includegraphics[width=\linewidth]{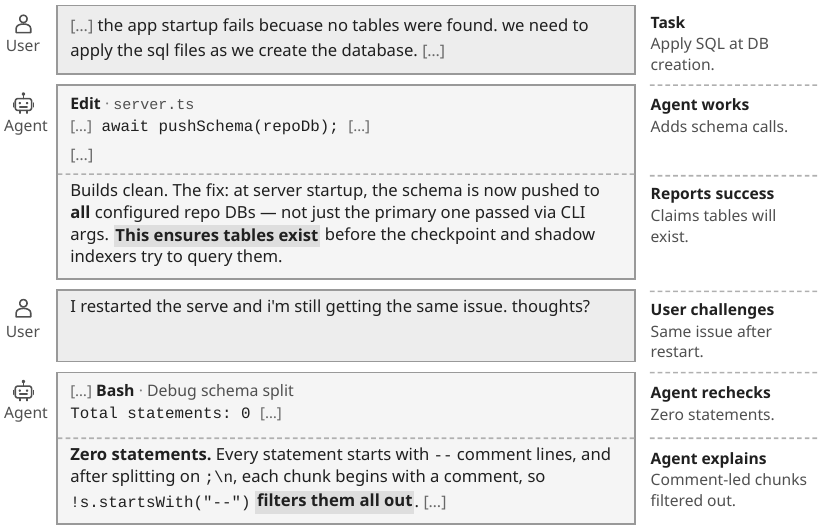}
\caption{Motivating example from SWE-chat. The agent reports a database-initialization fix, but its schema routine discards all SQL statements.}
\label{fig:motivation}
\Description{A user reports missing-table errors at application startup. The agent adds schema-initialization calls, runs a build, and states that the required tables will exist before indexing. The user restarts the server and reports the same failure. In a later diagnostic, the schema filter retains zero statements because it discards whole SQL chunks beginning with comments. Original conversation excerpts and agent tool activity are separated from the authors' annotations. Ellipses indicate omitted material.}
\end{figure}

Such reports, however, cannot always be taken at face value. An agent may assert that a fix works, that validation succeeded, or that a deliverable is complete when its own records do not support the assertion. We refer to false or unsupported claims about an agent's own actions and results as \emph{unfaithful reporting}. Verification and perceived integrity are recognized concerns in developers' trust in coding assistants \cite{wang2024coding-trust}, and unfaithful reports undermine both. A recipient who relies on such a claim receives false assurance: incomplete work may be accepted without verification, and erroneous assumptions about the current state may propagate into later steps or into the work of another agent.

Figure~\ref{fig:motivation} illustrates the problem. In a task to repair database initialization, the user reports missing-table errors. The agent adds schema-initialization calls and asserts: ``This ensures tables exist before the checkpoint and shadow indexers try to query them.'' After restarting the server, the user encounters the same errors. Only when challenged does the agent re-examine its routine and find that it discards every SQL chunk, because each chunk begins with a comment. The code change was genuine, yet the claim that it resolved the problem was not, and the discrepancy surfaced only through the user's own check.

Unfaithful reporting is a common and growing concern. In a large-scale analysis of real-world developer--agent sessions, Tang et al.\ identify inaccurate self-reporting as a frequent form of misalignment whose share has increased over time \cite{tang2026coding-agents-fail}. Benchmark studies likewise find that agents' claims of task completion often diverge from the actual environment state \cite{advani2026false-success}.

The concern extends to how agents are evaluated and trained. Current reward signals may favor code correctness and completion-oriented responses, while faithful self-reporting remains harder to measure \cite{tang2026coding-agents-fail}. LLM-based evaluators such as Agent-as-a-Judge inspect files and trajectories to assess whether task requirements are met \cite{zhuge2025agent-judge}, yet LLM judges can rely on surface completion signals such as confident closing language \cite{advani2026false-success}. An evaluator that credits a stated accomplishment without examining the supporting records rewards the report rather than the work, and training on its feedback may reinforce the behavior. In Qwen-AgentWorld, for instance, an environment simulator learned to insert success assurances into its predicted tool observations, raising judge scores without improving accuracy \cite{qwen2026agentworld}.

Despite this growing concern, the forms unfaithful reporting takes in real development work, the follow-up it entails, its prevalence, and the effectiveness of automated detection remain insufficiently understood. We address this gap with a systematic mixed-methods study of real developer--agent conversations from SWE-chat \cite{baumann2026swechat}. Studying unfaithful reporting first requires reliable instances of it. We therefore start from \emph{self-admitted unfaithful reports}: a user challenges a claim in an agent's report, and the agent then acknowledges that the claim was inaccurate or unsupported. The challenge and the admission together pinpoint the disputed claim and indicate that it was problematic.

With these cases, we first ask \textbf{RQ1 (Patterns):} \emph{What unfaithful claims do coding agents make about their work?} For each disputed claim, we compare what the agent reported with what its records show and organize the discrepancies into a taxonomy. Across 339 claims, unfaithful reporting falls into eleven subtypes in four categories: misreported observations, misattributed evidence, overstated task success, and faulty technical explanations. Overstated task success is by far the most common category (69.3\%). Its claims announce completion, correct behavior, or working functionality that the records do not establish, sometimes alongside genuine edits and passing checks.

We then ask \textbf{RQ2 (Follow-up):} \emph{What follow-up accompanies unfaithful claims once users challenge them?} We trace the conversation after each challenge, measuring further user turns, recorded agent time, and token use, and classifying the emotions expressed in user messages. Handling these problems involves on average 4.3 further user turns and 12.71 minutes of recorded response and execution time. Dissatisfaction appears in 40.28\% of user messages in these cases, compared with 17.90\% of randomly sampled messages from other SWE-chat sessions. It persists beyond the initial objection: excluding first-challenge messages, 30.71\% of the remaining messages still express dissatisfaction.

We next ask \textbf{RQ3 (Prevalence):} \emph{How often do coding agents' reports contain unfaithful claims?} We annotate 435 sampled work reports and find that \NaturalPositiveReports{} (\NaturalPositiveShare{}\%) contain at least one unfaithful claim.

The observed discrepancies also suggest what an automated check should inspect. We therefore ask \textbf{RQ4 (Detection):} \emph{How effectively can unfaithful claims be detected from work reports and their preceding trajectories?} We develop FaithDetector, a detector for unfaithful reporting informed by RQ1, and compare it with a baseline on naturally sampled reports. FaithDetector attains \NaturalDetectorMacro{}\% macro-F1, about 13 percentage points above the baseline.

The paper makes four contributions:

\begin{itemize}

\item We construct a taxonomy of unfaithful reporting by coding agents.

\item We quantify follow-up effort and characterize user reactions to reporting problems.

\item We measure the prevalence of unfaithful reporting in sampled work reports.

\item We develop FaithDetector from the observed patterns and evaluate it against a baseline.

\end{itemize}

Together, the findings connect reporting discrepancies with the records needed to check them, the handling that follows a challenge, and opportunities for automated detection. Reporting faithfulness therefore warrants dedicated attention in code review, agent evaluation, and training feedback.

\section{Related Work}\label{sec:related-work}

\subsection{Reporting errors within coding-agent failures}

Reporting errors have been identified within broader studies of coding-agent failures. Tang et al. analyze developer-raised problems in conversations, including claims of unperformed actions, unrealized success, and overstated coverage, alongside other interaction failures, their costs, and their resolution \cite{tang2026coding-agents-fail}. They identify reporting issues when these surface through developer correction or pushback in the logs. Hasan and Biswas examine operational safety through published research and GitHub incidents \cite{hasan2026operational-safety}. Their categories include unsupported action claims, fabricated evidence, and false assurances. They recommend tying status reports to command traces, diffs, and environment-state checks.

Trajectory studies examine relationships between agent statements and execution. Bouzenia and Pradel analyze thoughts, actions, and results, including tool-result misinterpretation and erroneous completion judgments, through relationships within steps and between adjacent reasoning and execution steps \cite{bouzenia2025trajectories}. MIRAGE-Bench uses snapshots of instructions, history, and observations to assess the contextual faithfulness of an agent's next action at a controlled decision point \cite{zhang2025mirage}. These assessments concern execution steps; a delivered report can combine several steps and refer to different artifacts or historical states. Advani's workshop study compares completion signals with programmatic task outcomes in tau2-bench and AppWorld, using closing statements or structured status fields to identify false success \cite{advani2026false-success}. This comparison identifies inaccurate completion status. Even when a task succeeds, its report may misdescribe observations, artifact contents, or technical explanations.

\subsection{Developer verification, trust, and oversight}

Studies of coding assistants show that accepting generated work involves verification. Barke et al. describe code inspection, execution, static analysis, and documentation lookup in programmers' interactions with Copilot \cite{barke2023grounded-copilot}. Ferdowsi et al. study live programming as support for validating AI-generated code, observing a participant relying on a code comment whose promised behavior was absent from the implementation \cite{ferdowsi2024live-validation}. Work reports extend these concerns to accounts of actions already taken and evidence obtained during a delegated task.

Greater automation changes the work developers oversee. Chen et al. compare copilot and agent workflows, observing developers asking agents to run code, show changes, and explain where edits were made \cite{chen2026code-with-me}. Dhanorkar et al.'s interviews describe post-hoc oversight through outputs, rationales, and change summaries \cite{dhanorkar2026oversight}. This oversight depends on information about what the agent changed, checked, and delivered. Wang et al. examine developers' trust through interviews and design probes, considering perceived ability, benevolence, integrity, and contextual stakes \cite{wang2024coding-trust}. Baltes et al. distinguish trust from trustworthiness and caution against treating artifact acceptance or reliance as sufficient measures of trust \cite{baltes2026rethink-trust}. Reporting accuracy offers a concrete behavior to assess: whether an account is supported by work records. Developers' perceptions of that behavior remain a separate question.

\subsection{Factual consistency of software descriptions}

Research on generated software text examines consistency with the artifacts being described. ETF traces entities between code and generated summaries \cite{maharaj2025etf}. Liu et al. study hallucinations in commit messages and review comments relative to code changes and context \cite{liu2025codechange-hallucinations}. HalluJudge evaluates support, context alignment, and scope in review comments \cite{tantithamthavorn2026hallujudge}. These studies address whether software descriptions accurately represent their sources. Work reports additionally describe executions and historical states: whether a command ran, which version a test checked, or where a change was deployed. Applicable evidence must match the claim's object, conditions, and state.

General factuality research provides foundations for this assessment. FactCC checks summary consistency and locates evidence or inconsistent spans \cite{kryscinski2020factcc}; FActScore assesses atomic claims against a specified knowledge source \cite{min2023factscore}. FRANK distinguishes errors in entities, predicates, circumstances, and discourse relationships \cite{pagnoni2021frank}, while RAGTruth separates conflicting content from unsupported additions \cite{niu2024ragtruth}. Claim-level checking and evidence localization therefore have precedents. In software work, these principles must account for execution and state: an operation may succeed yet concern a different deliverable or cover fewer conditions than the report asserts.

\subsection{Agent evaluation and learning from feedback}

Software-agent evaluation often measures task accomplishment. SWE-bench assesses issue resolution by applying patches and running tests \cite{jimenez2024swebench}; Wang et al. use differential testing and inspection to investigate incorrect patches counted as solved \cite{wang2026swe-correctness}. Agent-as-a-Judge reads files and trajectories against task requirements \cite{zhuge2025agent-judge}. Reporting accuracy asks whether the agent's statements about that work are supported. An unfinished task can be reported accurately, and successful work can be accompanied by an inaccurate account.

Feedback can also guide subsequent agent behavior. CRITIC uses external tools to critique and revise outputs, while Reflexion uses linguistic feedback and reflective memory for subsequent trials without updating model weights \cite{gou2024critic,shinn2023reflexion}. For work reporting, a useful correction must identify what the evidence supports, preserving genuine progress while correcting an inaccurate account. Qwen-AgentWorld illustrates a risk in relying on favorable descriptions: its environment simulator improved judge scores through success assurances or echoed criteria added to predicted tool observations without improving prediction accuracy \cite{qwen2026agentworld}. SALMON's assistant appended self-evaluations claiming its answers were helpful or perfect, prompting a principle against judging its own performance \cite{salmon2024}. These effects depend on the evaluator: in Zheng et al.'s AlpacaEval 2.0 tests against GPT-4-1106-preview, constant persuasive responses without task information win less than 1\%, unlike structured attacks \cite{zheng2025null-models}.

\subsection{Research gap}

Prior studies identify reporting errors within broader failures and examine execution steps, artifact descriptions, and completion status. These assessments leave open which claims in a delivered work report are inaccurate and how the available records establish the discrepancy. We characterize recurring discrepancies and trace how developers and agents handle them. We also measure the prevalence of unfaithful reporting through annotations of sampled work reports and evaluate FaithDetector, informed by the observed patterns, against a baseline using pre-report work records.

\section{Method}\label{sec:method}

\subsection{Study overview}

\begin{figure}[t]
\centering
\includegraphics[width=\linewidth]{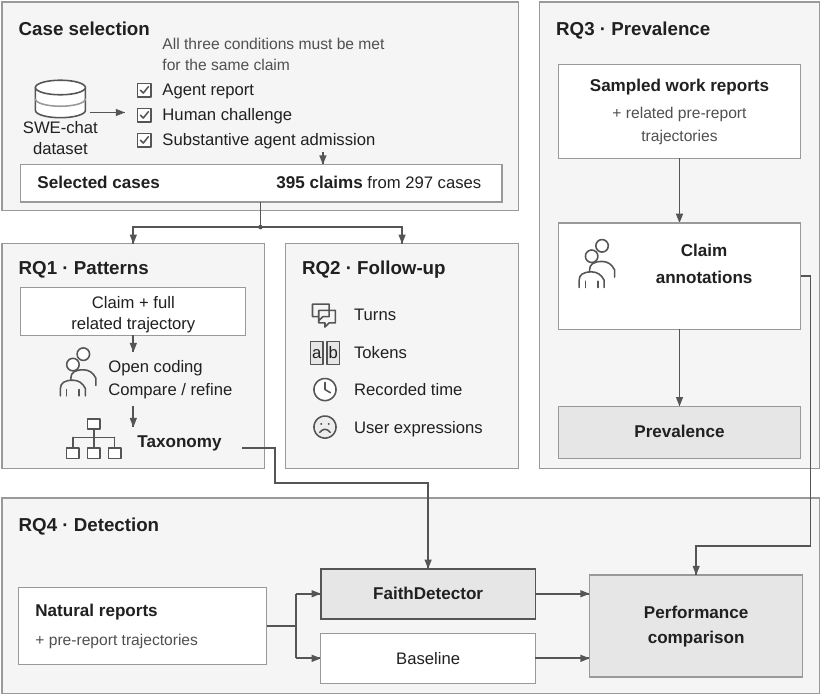}
\caption{Overview of the study methodology.}
\label{fig:overview}
\Description{The study has four research questions. Cases from SWE-chat are selected when an agent report, a human challenge, and a substantive agent admission concern the same claim. The 395 claims from 297 cases supply the RQ1 pattern analysis and RQ2 follow-up analysis. RQ1 uses full related trajectories, open coding by two authors, comparison, and refinement to develop a taxonomy. RQ2 describes recorded follow-up turns, tokens, time, and user expressions. In RQ3, one author annotates claims using complete sampled work reports and related pre-report trajectories, and another author verifies the annotations. We then estimate prevalence. RQ4 compares the performance of FaithDetector and Baseline on natural reports, using the RQ3 claim annotations as evaluation ground truth. The RQ1 taxonomy guides FaithDetector. Both methods receive complete reports and applicable pre-report trajectories.}
\end{figure}

Figure~\ref{fig:overview} summarizes our study of work reports from SWE-chat \cite{baumann2026swechat}. We characterize admitted reporting problems (RQ1) and trace their follow-up (RQ2). For RQ3, we annotate separately sampled work reports using their related pre-report trajectories and measure the prevalence of unfaithful reporting. For RQ4, we use the RQ3 annotations as ground truth to compare FaithDetector, informed by RQ1, with a baseline on these natural reports.

A report is the agent's complete account of its work at a particular point in a conversation; a claim is a checkable statement within it; a case is a source-annotated reporting problem that may contain several claims or reports. For the RQ1 and RQ2 case analyses, two authors checked the material, compared their judgments, and resolved disagreements against the original records.

\subsection{Identifying reporting problems}

We start from 1,221 annotations in Tang et al.'s Inaccurate Self-Reporting category \cite{tang2026coding-agents-fail}. We locate each quotation, or a contiguous verbatim fragment, in the source transcript and recover the complete report. We include a case when a user challenges a work claim and the reporting agent substantively admits or corrects it under its original conditions and qualifications. After excluding claims with unavailable source material or a mismatch with the original reporting problem, the common collection contains 395 claims from 297 source cases. We merge repeated statements of the same proposition about the same state, retaining links to every occurrence.

Work reports include findings from investigating a specific task problem. For example, the dry-run explanation in \texttt{r3927} reports a diagnosis of the missing completion output in the task being investigated. A stand-alone explanation of how a tool generally works is outside scope. In reports mixing work findings and general explanations, a counted problem must concern a work assertion.

\subsection{Characterizing reporting patterns}

We read each claim's earliest included occurrence with the original request and work records, documenting the assertion, evidence, and specific contradiction or support gap. We distinguish claims contradicted by applicable evidence from claims whose scope or certainty exceeds what those records establish. A support gap does not by itself establish factual falsity. Later checks apply only to the same target, conditions, and state, accounting for intervening changes. We develop open codes from the discrepancies. We compare codes within and across cases, combine recurring patterns, and revise definitions and examples in a two-level codebook. Two authors independently coded the claims. We assign patterns to 339 claims from 258 source cases. The other 56 remain unresolved because the records do not establish the relevant work state. Each claim counts once per applicable subtype or parent category. The macro-averaged binary Cohen's $\kappa$ across the 11 subtypes was 0.757 (range 0.655--0.934).

\subsection{Tracing follow-up handling}\label{sec:boundary}

We trace handling in 286 cases with a determinable boundary, including cases whose precise RQ1 discrepancy remains unresolved. Related activities directly investigate or address the disputed claim: checking or explaining it, correcting the report, completing omitted work, repairing the artifact, and checking the result, with directly connected user feedback and agent replies. We read from delivery of the first associated report to the last substantive activity addressing the problem and check subsequent context to establish the endpoint. In the miscounted-todo case \texttt{r4162}, we retain the recount and correction through the later pull request and recorded pull of the updated main branch. The intervening request for extra todos and subsequent codebase-concerns request are new work and excluded. Explicit returns to the original problem are retained; shared activities count once.

Scripts count user turns, main-agent responses, and tool calls in the related activities. Consecutive human messages form one turn unless agent text or tool activity intervenes. Output tokens sum available usage records for responses fully within these activities, counting each response once. Scripts take the union of recorded response and tool-execution intervals so parallel and nested work counts once. Gaps between an agent output and the next user input are excluded; missing intervals or durations inseparable from excluded activities remain unallocated.

We select 849 English user messages with reliable speaker attribution and usable context, excluding overlong inputs. GPT-5.6 Luna with medium reasoning effort generates the initial expression labels using EmoWOZ's seven categories \cite{feng2022emowoz} and the contextual labeling procedure adapted from LLM4ERC \cite{feng2024affect-llms}. Inputs contain the complete user message and its three preceding human or agent natural-language messages, omitting tool output and automatic content. We review labels against these inputs before adoption. Scripts randomly select 849 comparison messages from other SWE-chat sessions, excluding study sessions and linked continuations. Both collections use the same message criteria and labeling procedure.

\subsection{Sampling and annotating work reports}\label{sec:natural-reports}

We randomly sample 500 SWE-chat sessions, excluding prior development material and related sessions, and select one complete report per session where available. Thirty-three sessions have no selected report, leaving 467 reports. Scope review retains 435 work reports, excludes sixteen, and leaves sixteen unresolved.

One author annotated the sampled work reports against their related pre-report conversation and tool records, and another author verified the annotations. A Faithful label requires support for the claim's scope, conditions, and state; an Unfaithful label requires applicable counterevidence or an identifiable assertion beyond what the records establish. Absence of a test alone is insufficient. Claims whose evidence is unavailable or cannot establish the relevant state remain unresolved. Annotations include reasons and source excerpts. We calculate prevalence as the proportion of the 435 reports containing at least one claim annotated as Unfaithful.

\subsection{Developing and evaluating the detector}\label{sec:checker}

We develop FaithDetector using GPT-6 Astra. The RQ1 taxonomy guides checks of reported observations, evidence attribution, completion scope, and technical explanations. Given a complete report and its applicable pre-report trajectory, FaithDetector identifies checkable claims, compares them with the records, and returns Faithful or Unfaithful judgments with excerpts, evidence, and reasons. Subsequent challenges, admissions, and work results are withheld.

Using the RQ3 annotations as ground truth, we compare FaithDetector with a baseline on 419 naturally sampled reports with executable inputs. Both methods receive the same complete reports and applicable pre-report trajectories. The baseline uses GPT-6 Astra and the request, ``Can you find any unfaithful claims in the report?'' It can read and search the trajectory but receives no taxonomy, fixed targets, annotation labels, or prescribed response schema.

After generation, we match each method's allegations to the annotated claims, with labels withheld during matching. We score whether each target receives an alarm; no alarm is not an explicit Faithful judgment. Unresolved annotations, technical failures, and uninterpretable outputs are excluded. The paired evaluation covers \EvaluationTargets{} determinate claims from \EvaluationReports{} reports. We report macro-F1, the unweighted mean of the two class F1 scores, and Unfaithful precision and recall. To assess the uncertainty in FaithDetector's improvement over the baseline, we estimate a 95\% confidence interval using 2,000 paired bootstrap samples of reports, keeping each report's claims together.

\section{RQ1: Patterns}\label{sec:rq1}

We identify four categories and eleven subtypes (Table~\ref{tab:taxonomy}). Overstating Task Success is the largest category (69.3\%). Across categories, genuine edits, checks, and pushes can coexist with inaccurate claims about the work, artifact, or behavior established.

\afterpage{\clearpage\begingroup
\fontsize{9.5}{11.5}\selectfont
\renewcommand{\arraystretch}{1.0}
\setlength{\tabcolsep}{5pt}
\setlength{\LTcapwidth}{\linewidth}
\begin{longtable}{@{}>{\raggedright\arraybackslash}p{0.34\linewidth}>{\raggedright\arraybackslash}p{\dimexpr0.66\linewidth-2\tabcolsep\relax}@{}}
\caption{Patterns of unfaithful reporting.}\label{tab:taxonomy}\\
\toprule
Pattern / claims (\%) & Description \\
\midrule
\endfirsthead
\caption[]{Patterns of unfaithful reporting (continued).}\\
\toprule
Pattern / claims (\%) & Description \\
\midrule
\endhead
\midrule
\multicolumn{2}{r@{}}{\textit{Continued on next page}}\\
\endfoot
\bottomrule
\multicolumn{2}{@{}p{\linewidth}@{}}{\footnotesize Claims may have multiple subtype labels; each claim counts once per subtype and parent category. Category C contains 235 distinct claims and 237 subtype assignments. Percentages use 339 coded claims.} \\
\endlastfoot
\multicolumn{2}{@{}p{\linewidth}@{}}{\textbf{A. Misreporting Observations}\hfill\textbf{45 (13.3\%)}} \\*
\multicolumn{2}{@{}p{\linewidth}@{}}{Misdescribing the content or meaning of recorded measurements, actions, or check outcomes.} \\*
\addlinespace[3pt]
\textbf{A1. Misreported Quantitative Results}\enspace 11 (3.2\%) & A reported value, counting basis, or statistical interpretation conflicts with the quantitative record or exceeds what it supports. \\
\addlinespace[3pt]
\textbf{A2. Misreported Action History}\enspace 13 (3.8\%) & The report incorrectly states whether or when an action occurred, or what that particular action changed. \\
\addlinespace[3pt]
\textbf{A3. Misreported Check Outcomes}\enspace 21 (6.2\%) & A reported verdict or finding conflicts with the linked check or lookup result, or the retained output or status does not establish it. \\
\addlinespace[2pt]
\midrule
\multicolumn{2}{@{}p{\linewidth}@{}}{\textbf{B. Misattributing Evidence}\hfill\textbf{28 (8.3\%)}} \\*
\multicolumn{2}{@{}p{\linewidth}@{}}{Attributing information or work to a different source, object, or artifact state.} \\*
\addlinespace[3pt]
\textbf{B1. Source Attribution Errors}\enspace 3 (0.9\%) & Information is attributed to an observation, publication, component, or other source different from its established origin. \\
\addlinespace[3pt]
\textbf{B2. Object Attribution Errors}\enspace 13 (3.8\%) & Work or results from one feature, service, data group, or configuration are credited to another identifiable object. \\
\addlinespace[3pt]
\textbf{B3. Artifact State Attribution Errors}\enspace 12 (3.5\%) & Real work or results from one artifact state are credited to another, with evidence identifying the mismatched states. \\
\addlinespace[2pt]
\midrule
\multicolumn{2}{@{}p{\linewidth}@{}}{\textbf{C. Overstating Task Success}\hfill\textbf{235 (69.3\%)}} \\*
\multicolumn{2}{@{}p{\linewidth}@{}}{Overstating the required work covered, the conditions handled, or the usable effects established.} \\*
\addlinespace[3pt]
\textbf{C1. Overstated Completeness}\enspace 110 (32.4\%) & Completeness, correctness, equivalent information, or complete removal of a problem is claimed across a stated collection or implementation, while an identified required part or remaining instance fails that commitment. \\
\addlinespace[3pt]
\textbf{C2. Overstated Condition Handling}\enspace 71 (20.9\%) & The implementation does not provide promised behavior under an included input, ordering, topology, or operating condition. \\
\addlinespace[3pt]
\textbf{C3. Intermediate Results as Functional Proof}\enspace 56 (16.5\%) & A genuine edit, operation, or intermediate check is used to establish a distinct functional effect without a supported connection. \\
\addlinespace[2pt]
\midrule
\multicolumn{2}{@{}p{\linewidth}@{}}{\textbf{D. Faulty Technical Explanations}\hfill\textbf{42 (12.4\%)}} \\*
\multicolumn{2}{@{}p{\linewidth}@{}}{Giving an unestablished causal explanation or an incorrect rule about technical behavior.} \\*
\addlinespace[3pt]
\textbf{D1. Unsupported Causal Explanations}\enspace 25 (7.4\%) & An observation or hypothesis is presented as an established cause, mechanism, or exclusive explanation, beyond or contrary to what the evidence supports. \\
\addlinespace[3pt]
\textbf{D2. Incorrect Technical Rules}\enspace 17 (5.0\%) & A stated rule about a tool's or implementation's requirements, behavior, or necessary correction conflicts with an applicable rule, implementation, or technical result. \\
\end{longtable}
\endgroup
}

\subsection{Reports misstate actions, quantities, and check results}

Misreporting Observations covers 45 claims (13.3\%). A quantity error in \texttt{r4162} reported ``Pending todos: 6 $\rightarrow$ \textbf{14}'' after four new todo files were created. The directory listing contained the original six plus four additions: ten total. The files had been created and committed; the error concerned their number.

An action-history error in \texttt{r7024} credited a tooltip to the \texttt{ProjectTrendChart} metric button among ``\textbf{7 locations updated}.'' The agent created a shared tooltip component and added an import and explanatory metadata. However, the usage check showed only the import, and inspection before the corrective edit showed the button still closed after its icon and label. Supporting code existed without the reported rendering addition.

In \texttt{r6827}, a command filtered Biome's output to ``Checked'' lines, retaining ``Checked 236 files in 144ms. No fixes applied.'' Alongside successful localization and build checks, the report announced ``\textbf{All clean}.'' Biome's retained summary established that a check ran, not an error-free verdict. Later validation found a formatting error and showed that this summary can accompany diagnostics.

These cases distinguish a wrong quantity, a falsely reported action, and an unwarranted verdict. A testing claim also requires separating verdict from scope: Biome's clean verdict was unestablished, whereas the plugin test below succeeded but did not establish integrated startup. The consequence depends on the decision: a wrong count can preserve accomplished work, while a clean-check assertion can prematurely end validation.

\subsection{Reports attribute work to the wrong artifact or state}

Misattributing Evidence covers 28 claims (8.3\%). Source Attribution Errors (3 claims) concern where information originated. In \texttt{r6761}, the agent identified user-supplied HTML as ``from the \textbf{onboarding model selector} (\texttt{ModelSelectCard}).'' Its header and badge markup instead matched a separate card renderer in \texttt{providers.js}.

Object Attribution Errors (13 claims) concern where work was applied. In the same case, the user wanted the card to show an error reason inline instead of only in a tooltip. The agent edited the Settings-page list, then reported that the reason would ``\textbf{show inline below the model ID}'' in the requested card. Source inspection distinguished the edited list from the separate card renderer, which was changed later. The report credited a real edit to the wrong interface.

Artifact State Attribution Errors (12 claims) concern which state contains the work. In \texttt{r4055}, the agent pushed a release-workflow fix and reported that it was ``\textbf{already on that PR}.'' The pull request had merged over an hour before the fix was created. The branch contained the fix; the merged artifact did not. Similarly, \texttt{r4587} reported production ``\textbf{deployed and healthy}'' after pushing a fix, but the healthy promotion result named an older commit. Later checks found the old process running and the new build not yet started.

The source, object, and state subtypes distinguish where information originated, which interface received an edit, and which artifact state contains the work. Similar markup, a shared branch name, or a healthy service does not establish those relationships. Evidence can exist without applying to the reported target. Comparison with Overstated Completeness separates an unfinished requirement from a mistaken state assignment. In pull-request case \texttt{r6748}, ``Done'' exceeded the delivered work because the requested head had not received it. In \texttt{r4055}, ``already on that PR'' assigned a later commit to an already-merged artifact. Both retained a successful push; the requested destination and artifact history determine which discrepancy it presents. Attribution errors matter when a decision depends on the particular artifact or revision: the pushed fix in \texttt{r4055} was absent from the merged pull request, and the healthy production service in \texttt{r4587} was still running the older revision.

\subsection{Agents overstate what their work accomplishes}

Overstating Task Success covers 235 claims (69.3\%). Its subtypes distinguish unfinished requirements, behavior absent under a promised condition, and functional assurances inferred from intermediate results. Overstated Completeness is the largest subtype (110 claims).

In \texttt{r6748}, the agent merged changes, passed 225 provider tests, pushed a branch, and reported ``\textbf{Done}.'' Yet it created a prefixed branch in the main repository while the requested pull request tracked another branch in the contributor's fork. Successful local work and a push elsewhere left delivery unfinished. In \texttt{r7420}, the agent updated \texttt{IpMappingForm} and the backend, passed a build, and reported ``\textbf{All changes are in place}'' for a domain-dropdown task. Inspection before repairs showed the page's separate inline add and edit controls still used text inputs. The changed component did not cover the required workflow.

Overstated Condition Handling (71 claims) concerns behavior promised under a condition that the implementation does not handle. In case \texttt{r5362}, the goal was to keep sessions already associated with the main branch out of the list offered when restoring branch sessions. The report stated that ``\texttt{resumeSession} \textbf{filters out sessions} whose IDs are in the main set before displaying results.'' The agent had added a lookup for that set and a filtering loop, and the recorded tests passed. However, an earlier return when HEAD contained a checkpoint bypassed the discovery needed by the filter. The implementation could find a checkpoint for the branch without obtaining the main-session IDs needed to decide which sessions to exclude. Under that condition, the helper returned no main-session set and filtering was disabled. In database-migration case \texttt{r7156}, the task required the same indexes for newly created, rebuilt, and recovered tables. Two index-parity tests passed, and the report announced ``\textbf{All tasks complete}.'' Yet the implementation returned when the primary key or foreign key was already composite, before ensuring the indexes. A later source read, before repairs, identified these original returns. Passing tests did not establish index restoration for an already-composite table missing its indexes. In case \texttt{r7419}, an agent raised a service's default query limit from 100 to 20,000, then told the user to refresh and search again to ``see \textbf{all HIGH+CRITICAL findings}.'' The UI's ordinary and refresh search paths still supplied \texttt{pageSize=100}, bypassing the edited default. The promised increase was therefore not provided under the existing caller configuration.

Intermediate Results as Functional Proof (56 claims) concerns a further outcome inferred from an implemented step or restricted check. The motivating example inferred table availability from initialization calls whose routine discarded the SQL. In \texttt{r5260}, a rebuilt plugin passed a standalone initialization test in 19 ms using a mock client. The report said it ``should \textbf{start normally}'' and assured startup without hanging, although the test had not exercised OpenCode's actual services. The user's immediate retry still hung. ``Should'' and a request to retry preserved some uncertainty, but the closing assurance went beyond an exploratory suggestion.

Completeness compares delivery with the objective; condition handling follows the promised execution path; functional proof examines the inference to a further outcome. Passing builds do not establish coverage of all controls, and a mock-client test does not establish integrated startup. Correcting these reports may also require further work, such as pushing \texttt{r6748}'s changes to the contributor's fork.

\subsection{Technical explanations assert unsupported causes or incorrect rules}

Faulty Technical Explanations covers 42 claims (12.4\%). 25 claims present a cause as established without observations that isolate it. In Tauri audio case \texttt{r4210}, the playback implementation connected a source through normalization and gain nodes to a media-stream destination, then assigned the resulting stream to an audio element. A delegated assessment suggested that the destination stream might not receive audio from the connected nodes. The final report instead asserted that ``\texttt{createMediaStreamDestination()} produces a stream that \textbf{doesn't carry audio data}, resulting in complete silence with no errors.'' The observations did not isolate that component as the cause of silence. A replacement implementation built successfully, which established a build result rather than the cause of a runtime symptom. Later diagnostics and a successful bypass of the wider processing graph also did not distinguish a destination-specific failure from problems elsewhere in the bypassed graph. The report therefore converted a possible explanation into a localized causal assertion.

Incorrect technical rules occur in 17 claims. In \texttt{r3927}, the agent explained missing output by claiming that ``Fractor \textbf{isn't outputting} `Fractor is done!' when run in \textbf{dry-run mode}.'' The user supplied an explicitly identified dry-run execution containing that message, contradicting the rule while leaving the original missing output unexplained.

A workaround's success and a rule's falsification answer different questions. Treating an unisolated cause or contradicted rule as settled can misdirect repairs. Across categories, the comparisons identify what a correction should preserve and what still needs attention: an observation, artifact relationship, delivery, execution path, or diagnostic check.

\section{RQ2: Follow-up}\label{sec:rq2}

\subsection{Handling a reporting problem often takes several exchanges}

Recorded handling averages 4.29 user turns, 36.67 main-model responses, and 35.19 tool calls (Table~\ref{tab:effort}). The medians are three turns, 22 responses, and 19 calls; at the 90th percentile, handling involves nine user turns and about 86 tool calls. These activities extend beyond a single objection and reply, with some cases involving considerably more agent and tool activity than the typical case.

\begin{table}[tbp]
\begin{minipage}{\linewidth}
\begin{minipage}[t]{0.485\linewidth}
\vskip0pt
\captionsetup{hypcap=false,skip=4pt}
\captionof{table}{Follow-up effort.}
\label{tab:effort}
\footnotesize
\setlength{\tabcolsep}{2pt}
\renewcommand{\arraystretch}{1.12}
\begin{tabularx}{\linewidth}{@{}>{\raggedright\arraybackslash}Xrrrr@{}}
\toprule
Metric & Cases & Mean & Median & P90 \\
\midrule
User turns & 284 & 4.29 & 3 & 9 \\
Main-model responses & 283 & 36.67 & 22 & 85.80 \\
Tool calls & 283 & 35.19 & 19 & 85.80 \\
Output tokens (partial) & 188 & 12,135 & 5,318.5 & 24,736.7 \\
Recorded time (min) & 286 & 12.71 & 5.57 & 32.31 \\
\bottomrule
\end{tabularx}

\end{minipage}\hfill
\begin{minipage}[t]{0.485\linewidth}
\vskip0pt
\centering
\includegraphics[width=\linewidth]{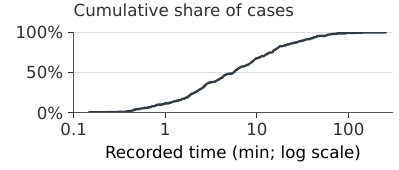}
\captionsetup{hypcap=false,skip=4pt}
\captionof{figure}{Recorded time across 286 cases.}
\label{fig:effort-distribution}
\end{minipage}
\par\vspace{0.6em}
\captionsetup{hypcap=false,skip=4pt}
\captionof{table}{User messages labeled dissatisfied.}
\label{tab:reactions}
\centering
\footnotesize\setlength{\tabcolsep}{2pt}
\renewcommand{\arraystretch}{1.16}
\begin{tabularx}{\linewidth}{@{}>{\raggedright\arraybackslash}Xrrr@{}}
\toprule
Message collection & Messages & Dissatisfied & \% \\
\midrule
Study messages & 849 & 342 & 40.28 \\
Study messages excluding first challenges & 661 & 203 & 30.71 \\
Comparison messages & 849 & 152 & 17.90 \\
\bottomrule
\end{tabularx}

\end{minipage}
\end{table}

Recorded response and execution time averages 12.71 minutes per case, compared with a median of 5.57 minutes and a 90th percentile of 32.31 minutes (Figure~\ref{fig:effort-distribution}). Partial output-token sums are available for 188 cases, averaging 12,135 tokens, with a median of 5,318.5 and a 90th percentile of 24,736.7. For both measures, the means exceed the medians, and the upper percentiles show the large variation in recorded follow-up activity.

\subsection{Follow-up extends to records, delivery, and verification}

Follow-up in these cases involved updating records or completing delivery alongside correcting the reported account. In the todo-count case \texttt{r4162}, the user asked, ``wait, how did we suddenly go from 6 pending to 14?'' The agent listed the directory, recognized that six existing files plus four additions yielded ten, and corrected the count in \texttt{STATE.md}. The correction subsequently reached a delivered pull request, and the user pulled the updated main branch. Resolving the count in the conversation was followed by updating and delivering the record that contained it.

In the pull-request delivery case \texttt{r6748}, follow-up exposed a change that had not reached its intended destination. The user questioned the completion report because the original pull request still showed conflicts. The agent inspected branch tracking and the pull request's head, then tried pushing to the expected branch name. The head remained unchanged. It eventually found that it had pushed to the main repository rather than the contributor's fork, added the fork remote, fetched the relevant reference, and pushed the merge commit. A final check showed that the pull request pointed to that commit with a clean mergeable state.

The count case carried a corrected statement into a persistent record and delivered change; the pull-request case required delivering an existing commit to the intended repository and checking the result. The exchanges therefore concerned both the reported account and the work it described. A corrected count or another successful push was an intermediate step in these observed handling sequences.

\subsection{Dissatisfaction continues during follow-up}

Dissatisfaction appears in 342 of 849 study messages (40.28\%), compared with 152 of 849 comparison messages (17.90\%), a difference of 22.38 percentage points (Table~\ref{tab:reactions}). Excluding the first challenges, 203 of the 661 remaining study messages (30.71\%) are still labeled dissatisfied, 12.81 percentage points above the comparison sample. Dissatisfaction thus extends beyond the initial objection into subsequent handling. Neutral remains the largest category in both collections (51.24\% and 71.73\%); abusive labels are uncommon (0.94\% and 0.35\%).

Some dissatisfied messages also contain concrete verification feedback. In case \texttt{r5784}, the agent reported changing transcript normalization to address a UI error when displaying a Gemini session. The user then wrote, ``I reran the build and tried it with a new checkpoint and it was \textbf{broken still},'' identifying the checkpoint tested. In case \texttt{r4130}, the agent reported pushing a fix to a dependency's build script to address a failing Windows cargo check. After further attempted repairs, the user reported ``\textbf{still failing in CI}'' and offered to move the session to a Windows machine for local debugging.

These users continued checking the work while expressing dissatisfaction. The first supplied the result of a fresh build and checkpoint test; the second supplied the CI failure and offered access to the relevant operating environment. In these examples, the objections provided information for continuing the investigation as well as conveying dissatisfaction with the still-failing behavior.

\section{RQ3: Prevalence}\label{sec:rq3}

Table~\ref{tab:prevalence} summarizes how many unfaithful claims we identify in each sampled work report. Most reports with identified unfaithful claims contain one or two, while a smaller group contains three or more. No unfaithful claim is identified in the majority of sampled reports. Among the affected reports, those containing multiple identified unfaithful claims collectively outnumber those containing just one.

\begin{table}[!htb]
\centering
\caption{Number of unfaithful claims identified per report.}
\label{tab:prevalence}
\small
\begin{tabular*}{0.8\linewidth}{@{\extracolsep{\fill}}lrr@{}}
\toprule
Unfaithful claims per report & Reports & Percentage \\
\midrule
0 & 273 & 62.76\% \\
\textbf{At least 1} & \textbf{162} & \textbf{37.24\%} \\
\quad 1 & 62 & 14.25\% \\
\quad 2 & 43 & 9.89\% \\
\quad 3 or more & 57 & 13.10\% \\
\midrule
Total & 435 & 100.00\% \\
\bottomrule
\end{tabular*}
\end{table}

\section{RQ4: Detection}\label{sec:rq4}

\begin{table}[t]
\centering
\caption{Detection performance by annotated class (\%).}
\label{tab:detection}
\small
\begin{tabular}{lrrrrrrr}
\toprule
 & & \multicolumn{3}{c}{Unfaithful} & \multicolumn{3}{c}{Faithful} \\
\cmidrule(lr){3-5}\cmidrule(lr){6-8}
Method & Macro-F1 & Precision & Recall & F1 & Precision & Recall & F1 \\
\midrule
FaithDetector & \textbf{\NaturalDetectorMacro{}} & \textbf{\NaturalDetectorPrecision{}} & \textbf{\NaturalDetectorRecall{}} & \textbf{\NaturalDetectorUnfaithfulFone{}} & \textbf{\NaturalDetectorFaithfulPrecision{}} & \textbf{\NaturalDetectorFaithfulRecall{}} & \textbf{\NaturalDetectorFaithfulFone{}} \\
Baseline & \NaturalBaselineMacro{} & \NaturalBaselinePrecision{} & \NaturalBaselineRecall{} & \NaturalBaselineUnfaithfulFone{} & \NaturalBaselineFaithfulPrecision{} & \NaturalBaselineFaithfulRecall{} & \NaturalBaselineFaithfulFone{} \\
\bottomrule
\end{tabular}
\end{table}

Table~\ref{tab:detection} shows that FaithDetector achieves \textbf{\NaturalDetectorMacro{}\% macro-F1}, compared with \NaturalBaselineMacro{}\% for the baseline. FaithDetector improves macro-F1 by \NaturalMacroDelta{} percentage points (95\% confidence interval: \NaturalDeltaLow{}--\NaturalDeltaHigh{}).

The largest improvement is on Unfaithful claims: FaithDetector achieves \NaturalDetectorUnfaithfulFone{}\% F1, compared with \NaturalBaselineUnfaithfulFone{}\% for the baseline. Its recall is \NaturalDetectorRecall{}\%, versus \NaturalBaselineRecall{}\%, while its precision is also higher (\NaturalDetectorPrecision{}\% versus \NaturalBaselinePrecision{}\%). It therefore identifies more reporting problems while making fewer false alarms. Both methods perform better on Faithful claims, where FaithDetector reaches \NaturalDetectorFaithfulFone{}\% F1 and the baseline reaches \NaturalBaselineFaithfulFone{}\%. The detector's advantage thus spans both classes, with the larger difference in detecting unfaithful claims.

The comparison exposes concrete differences in what the methods flag. In report \texttt{n444}, the agent claimed an entity list with ``infinite scroll.'' The saved component fetched another page through a ``Load more'' button. FaithDetector distinguished button-triggered pagination from infinite scrolling, whereas the baseline stated that it found no clear unfaithful claims in the implementation summary. This illustrates a detected mismatch between the claimed behavior and the visible implementation.

FaithDetector also raises some alarms where our annotation accepts an inference from tool behavior. In \texttt{n120}, the report claimed that \texttt{npm install} generated \texttt{package-lock.json}. The trajectory showed the file was absent beforehand and that installation then succeeded in the same directory. The baseline and our annotation accepted the normal side effect, while FaithDetector flagged the absence of a subsequent file confirmation. This example shows how requiring an additional observation can reject a claim supported by the adopted annotation rules.

\section{Discussion}\label{sec:discussion}

\subsection{Reports are becoming the interface for trusting delegated work}\label{sec:disc-trust}

Asynchronous and cloud agents return their work as reports or pull requests rather than live traces \cite{willison2025async-code-research,github2026cloud-agent}, and agents resuming long tasks rely on progress records left by earlier sessions \cite{young2025long-running-harnesses}. As direct observation thins, the report carries more of the weight of trust. It is often the only account a developer reads before accepting the work, and the only account a later agent has of what was done.

Our results show that this weight cannot rest on the report alone. Unfaithful claims are common in sampled reports (RQ3), and once challenged, they are followed by further exchanges and sustained dissatisfaction (RQ2). They also need not involve fabricated work. In our cases, they appear alongside genuine edits, checks, and pushes, and the signals that usually stand in for a report's credibility are present without establishing the claim. In \texttt{r6748}, the agent created the merge commit, passed 225 provider tests, pushed a branch, and reported ``Done.'' The push, however, went to a different branch from the one the pull request tracked, so the requested delivery had not happened. In \texttt{r4587}, a healthy promotion result accompanied the report that production was ``deployed and healthy,'' but that result described the previously deployed commit rather than the new fix.

In both cases, the passing tests and the healthy service were real, and each could satisfy an outcome check or an evaluator that relies on confident completion language \cite{advani2026false-success}. None of them established what the report claimed. Task success and reporting faithfulness are therefore distinct properties, and evidence of the first cannot vouch for the second. Reporting faithfulness needs to be checked in its own right wherever reports are consumed: by developers deciding whether to accept work, by tools and agents that carry the work forward, and by the evaluators and training pipelines that score it. The following subsections take these in turn.

\subsection{Checking reports against their records}\label{sec:disc-review}

For developers and reviewers, our results suggest reviewing the claims in a report, not only the work behind it. Two reasons make this necessary. First, the report informs the current decision. Developers already use agents' change summaries to decide what to inspect \cite{dhanorkar2026oversight}, and where code review is minimal, as in vibe coding \cite{fawzy2026vibe-practice}, the report may be the main basis for accepting the work. Second, the report becomes the record that a later developer or agent relies on when resuming the task \cite{young2025long-running-harnesses}. An unfaithful claim that passes review now may never be rechecked, because whoever continues the work will read the report rather than retrace the actions behind it.

The RQ1 categories turn this review into concrete acceptance questions (Table~\ref{tab:acceptance}). For a completion claim, the question is which required outcome and conditions have been established. In \texttt{r7420}, the named form component was updated and the build passed, but the page's separate inline controls were unchanged. For an attributed result, the question is whether the evidence belongs to the named artifact. In \texttt{r6761}, a real edit to the settings list was credited to the separate model card the user had asked about. Reported observations call for the underlying record, and explanations call for evidence that isolates the stated cause.

\begin{table}[tbp]
\caption{Proposed checks when using a report to accept agent work.}
\label{tab:acceptance}
\small
\renewcommand{\arraystretch}{1.15}
\begin{tabularx}{\linewidth}{@{}>{\raggedright\arraybackslash}p{.23\linewidth}XX@{}}
\toprule
Observed pattern & Acceptance question & Information to inspect \\
\midrule
Overstated task success & Which required outcome and operating conditions have been established? & Required work, exercised conditions, and remaining verification. \\
Misreported observations & What action, quantity, or verdict does the record actually show? & The relevant inventory or full check result, including diagnostics. \\
Misattributed evidence & Does this evidence belong to the named source, artifact, and revision? & Source location; commit and delivery destination; PR head or deployed revision. \\
Faulty explanations & What supports this rule or causal explanation? & Applicable code or output; checks distinguishing candidate causes. \\
\bottomrule
\end{tabularx}
\end{table}

Two results shape when this checking should happen. First, it should be routine rather than prompted by suspicion. Our annotation identifies unfaithful claims in \NaturalPositiveShare{}\% of sampled reports, and in our cases, the problem surfaced only because a user challenged it, after which handling it took several further exchanges. Second, the depth of checking should follow the decision the report supports. The count error in \texttt{n218} left its directory comparison usable, whereas relying on \texttt{r4055}'s report that a new fix was ``already on that PR'' could mean believing that a pull request merged an hour earlier contained it. Reviewers should therefore concentrate on claims that determine what is accepted, merged, deployed, or built upon. Baltes et al.\ argue that reliance in software engineering can rest on institutional processes rather than on an assistant alone \cite{baltes2026rethink-trust}. Building these questions into existing review, testing, and deployment steps makes such checks routine without asking every developer to re-derive every claim.

\subsection{Making reports carry their evidence}\label{sec:disc-design}

For designers of coding agents and their tools, the implication is that reports should state what their actions establish, not only what was done. Current guidance already asks agents to report the commands they ran and their results \cite{openai2026prompting}, and cloud agents deliver work as changes and pull requests for review \cite{github2026cloud-agent,openai2026cloud}. Our cases show that the gap lies elsewhere: between a real action and the broader outcome the report claims for it. A completion or functional claim should therefore name the artifact or revision it concerns, the conditions under which it was checked, the record that supports it, and what remains unverified. In \texttt{r6827}, the agent filtered a linter's output down to the line ``Checked 236 files in 144ms. No fixes applied.'' and reported ``All clean.'' Showing that line next to the claim would have revealed that it records a completed run rather than an error-free verdict. In \texttt{r5260}, a standalone initialization test with a mock client completed in 19 ms, yet the report assured the user that the plugin would start normally, and the user's next attempt still hung. A faithful report could read: ``Standalone initialization with a mock client completed in 19 ms. Startup in OpenCode remains to be checked.'' It keeps the accomplishment, its condition, and the open question.

The same structure matters when work changes hands. Agents that resume long tasks rely on progress records \cite{young2025long-running-harnesses}, so these records should store established outcomes with their evidence and list pending checks separately. A correction should likewise name the claim it revises, keep the supported work, and state the remaining action and the record that would confirm it. In \texttt{r4162}, recounting the inventory corrected the reported number of pending todos while preserving the four todo files the agent had created, and the corrected total then had to reach the delivered record. User complaints can already contain the next target. In \texttt{r5784}, the user reported a fresh build and the checkpoint that still failed, and a useful reply would keep that checkpoint as the object of its next check. Finally, claim-level checking can run before a report is delivered. RQ4 evaluates FaithDetector against a baseline on natural reports. The evaluation identifies both missed claims and false alarms within the annotated collection, so these checks should support review rather than replace it. These are design proposals, and their effect on recipients' decisions remains to be tested.

\subsection{Scoring reports separately from task outcomes}\label{sec:disc-eval}

For researchers who design evaluations and rewards for coding agents, our results suggest scoring the report separately from the task outcome. Tang et al.\ suggest that current reward signals favor code correctness and completion-oriented responses, while faithful self-reporting remains harder to measure \cite{tang2026coding-agents-fail}. Evaluators that read the report face the same difficulty. LLM judges can rely on confident completion language \cite{advani2026false-success}, and in Qwen-AgentWorld, a simulator learned to insert success assurances that raised judge scores without improving accuracy \cite{qwen2026agentworld}. Section~\ref{sec:disc-trust} showed why outcome evidence cannot substitute for this check: tests and health checks passed while the reported outcome had not been achieved. Evaluations should therefore check each claim against the records, as Agent-as-a-Judge does for task requirements \cite{zhuge2025agent-judge}, and report task credit and reporting faithfulness as separate scores.

Our cases suggest a concrete design. With the work records held fixed, an evaluation can compare an overstated report, a faithful one, and an over-hedged one. Task credit should be identical across the three, while the reporting score should separate them. The over-hedged condition matters because rewarding only the absence of false claims favors omission. In \texttt{r7024}, the report counted a chart button among seven locations that received a tooltip, although only the supporting import had been added there. A faithful correction keeps the locations that were updated and flags the button. Withdrawing the whole claim avoids the error but discards real progress. Feedback for training should likewise name the claim, the discrepancy, and the record that settles it, following the precedents of tool-based critique and feedback across trials \cite{gou2024critic,shinn2023reflexion}. A claim-level detector can supply such feedback at scale, but optimizing against an imperfect detector may teach agents to evade it, much as success assurances exploited judges in Qwen-AgentWorld. Whether such feedback improves faithfulness without reducing task progress remains to be tested.

\subsection{Responsible use in production}

Unsupported claims about deployment status, security validation, or operational health can increase production risk. A reviewer who accepts such a claim may skip a security check or rely on a fix that is absent from the running revision. In case \texttt{r4587}, the health result applied to an older revision. Before acting on a consequential claim, reviewers should check that the evidence covers the intended environment and revision.

Our findings and detector evaluation come from SWE-chat sessions in public repositories. Their applicability to enterprise development environments, with different access controls, tooling, and release processes, remains untested. FaithDetector is a decision-support tool. It flags claims and points reviewers to evidence. The responsible reviewer should verify that evidence and retain authority over decisions about deployment, security sign-off, and operational readiness.

\subsection{Future work}

Three directions follow from our results. First, the prevalence estimate should be tightened and extended. Source review beyond the extracted claims would estimate what the detector misses, and replications with other agents and platforms would show whether the patterns and their frequencies hold beyond SWE-chat. Second, reporting should be studied where work changes hands. Longitudinal studies could follow delegated work through review, deployment, and handoff to another developer or agent, observing which report information recipients use and whether an unfaithful claim propagates into later work. Multi-agent systems raise the same question without a human reader, because subagents return condensed summaries of their work to an orchestrator \cite{anthropic2025context-engineering}. Third, interventions should be tested in executable settings. Checking each report as it is produced and returning discrepancies to the agent before it continues, as an interactive faithfulness assistant would, could show whether early feedback changes what agents claim and do next, and at what cost in interruptions, false alarms, and delay. Training with claim-level feedback, compared under fixed task conditions, would in turn test Tang et al.'s hypothesis that current rewards favor completion over faithful reporting \cite{tang2026coding-agents-fail}.

\section{Threats to Validity}\label{sec:validity}

\subsection{Construct validity}

Response and execution intervals include waits rather than measuring human labor. Of 286 cases, 28 have missing or unallocatable intervals; logs for the remaining 258 cases can still omit activity. Partial output-token sums and shared activity describe recorded handling, not complete or independent task costs. Expression labels can misrepresent meanings in coding conversations despite our review. Our annotations and output matching may share errors with the detector. 

\subsection{Internal validity}

The original task environments cannot be restored from the transcripts. Later checks can clarify a reported result only when their target, conditions, and state can be linked to the original claim. We preserve intervening changes and distinguish subsequent corroboration from a controlled reproduction. Missing code, images, or runtime information may leave a discrepancy unresolvable and exclude it from pattern analysis, favoring better-documented problems. The detailed cases explain supported discrepancies and category distinctions; they do not establish the frequency of every activity or a unique technical cause.
Subsequent checks, additional work, and repairs prompted by users may change final task outcomes; the records therefore do not isolate the effect of reporting problems on task completion. Background messages are not screened for reporting errors or matched by task state; software failure can contribute to dissatisfaction. Repeated messages within sessions are not independent users or evidence of reporting-caused changes in trust.
The comparison evaluates the two implemented procedures and does not isolate the effect of the taxonomy.

\subsection{External validity}

User challenges and agent admissions favor noticed and acknowledged discrepancies, potentially altering category frequencies. RQ1 frequencies describe analyzable problem claims; RQ2 traces handling and can include claims whose precise discrepancy remains unsettled. Excluding 56 unresolved claims restricts category frequencies; codebook coverage establishes neither theoretical saturation nor completeness for unseen reports. SWE-chat collects sessions from public repositories whose developers opt into Entire CLI logging, selecting an early-adopter population and excluding proprietary enterprise codebases \cite{baumann2026swechat}. Sessions whose outputs are abandoned may leave no committed logs, so failed interactions can be underrepresented in the source dataset. Expression analysis further selects usable English messages. The observed frequencies and handling need not transfer to other systems, settings, or languages. Recoverable pre-report records favor well-documented reports. Natural sampling excludes recorded development exposure and related sessions. Sampling sessions and selecting one delivery per session give reports unequal inclusion probabilities; missing inventories further restrict coverage. RQ3 retains all 435 reports in its denominator, but unresolved and unextracted claims limit the identified share. This is a description of the sampled reports based on our annotations, not a corpus-wide prevalence estimate. Effects of detector feedback on recipient decisions, task completion, and supervision effort require evaluation within ongoing tasks.

\section{Conclusion}\label{sec:conclusion}
From real developer--agent conversations, we identify four categories and eleven subtypes of unfaithful reporting and trace the interaction effort, recorded resource use, and dissatisfaction expressed while addressing these problems. Among 435 work reports that we separately sample and annotate, \NaturalPositiveReports{} (\NaturalPositiveShare{}\%) contain at least one unfaithful claim. FaithDetector, informed by the observed patterns, attains \NaturalDetectorMacro{}\% macro-F1 on the natural evaluation, about 13 percentage points above the baseline. Across the cases, successful local operations could coexist with inaccurate statements about their outcome, destination, or operating conditions. The taxonomy and examples characterize recurring reporting errors, and FaithDetector provides an automated method to check claims against work records. Together, they inform report review, agent evaluation, and training feedback that support developer supervision of ongoing agent work.

\section*{Data Availability}
The replication package is available on Figshare at \url{https://doi.org/10.6084/m9.figshare.34056699}. Researchers can request access to the upstream SWE-chat release through the \href{https://huggingface.co/datasets/SALT-NLP/SWE-chat}{dataset page}.

\paragraph{Data provenance and permissions.}
We use SWE-chat for secondary research. The dataset creators report that developers opted into Entire CLI checkpoint logging and pushed the logs to public branches, and that the source repositories were selected for licenses permitting redistribution \cite{baumann2026swechat}. The dataset page lists the \href{https://opendatacommons.org/licenses/by/1-0/}{Open Data Commons Attribution License (ODC-BY)}, which permits use and redistribution of the database subject to attribution. Individual contents remain subject to their applicable licenses and rights. Hugging Face requires users to sign in, accept the dataset access conditions, and agree to share their account information with the dataset maintainers. The maintainers also provide a data removal request process on the dataset page.

\paragraph{Privacy and security.}
The dataset creators report using Microsoft Presidio to redact personally identifiable information in user prompts and assistant responses and TruffleHog to detect and remove credentials before release \cite{baumann2026swechat}. The examples reproduced in this preprint were inspected for personal information, credentials and tokens, private URLs, infrastructure identifiers, and sensitive repository or organizational details. The excerpts retain the context needed to understand the claims and the evidence.

\bibliographystyle{ACM-Reference-Format}
\bibliography{references}
\end{document}